\documentclass[aps,prl,twocolumn,superscriptaddress,]{revtex4-1}
\usepackage{amsmath}
\usepackage{amsfonts}
\usepackage{graphicx}
\usepackage{color}
\usepackage{dsfont}
\usepackage{verbatim}
\usepackage{mathtools}
\usepackage[normalem]{ulem}
\usepackage{comment}
\usepackage{stackrel}
\usepackage{bm}
\usepackage[pdftex]{hyperref}
\hypersetup{colorlinks = true, urlcolor = black, linkcolor = black, citecolor = black}

\definecolor{myblue}{rgb}{.93, .93, 1}

\newcommand{\bsub}{\begin{subequations}}
	\newcommand{\esub}{\end{subequations}}

\newcommand{\vex}[1]{\bm{\mathrm{#1}}}

\begin{document}
	
\title{Acoustic-phonon-mediated superconductivity in rhombohedral trilayer graphene}
	
\author{Yang-Zhi~Chou}\email{yzchou@umd.edu}
\affiliation{Condensed Matter Theory Center and Joint Quantum Institute, Department of Physics, University of Maryland, College Park, Maryland 20742, USA}
	
\author{Fengcheng~Wu}\email{wufcheng@whu.edu.cn}
\affiliation{School of Physics and Technology, Wuhan University, Wuhan 430072, China}
	
\author{Jay D. Sau}
\affiliation{Condensed Matter Theory Center and Joint Quantum Institute, Department of Physics, University of Maryland, College Park, Maryland 20742, USA}
	
\author{Sankar Das~Sarma}
\affiliation{Condensed Matter Theory Center and Joint Quantum Institute, Department of Physics, University of Maryland, College Park, Maryland 20742, USA}	
\date{\today}
	
\begin{abstract}
Motivated by the observation of two distinct superconducting phases in the moir\'eless ABC-stacked rhombohedral trilayer graphene, we investigate the electron-acoustic-phonon coupling as a possible pairing mechanism. We predict the existence of superconductivity with the highest $T_c\sim 3$K near the Van Hove singularity. Away from the Van Hove singularity, $T_c$ remains finite in a wide range of doping.
In our model, the $s$-wave spin-singlet and $f$-wave spin-triplet pairings yield the same $T_c$, while other pairing states have negligible $T_c$. Our theory provides a simple explanation for the two distinct superconducting phases in the experiment and suggests that superconductivity and other interaction-driven phases (e.g., ferromagnetism) can have different origins.
\end{abstract}
	
	\maketitle
	
\textit{Introduction. --} Discovery of correlated insulators and  superconductivity in the magic-angle twisted bilayer graphene \cite{tbg1,tbg2} initiates the exploration for exotic phenomena in moir\'e systems \cite{Yankowitz2019,Polshyn2019,Cao2020PRL,Sharpe2019,Lu2019,Kerelsky2019,Jiang2019,Xie2019_spectroscopic,Choi2019,Serlin2020,Park2021flavour,Chen2019signatures,Burg2019,Shen2020correlated,Cao2020tunable,Liu2020tunable,Park2021tunable,Hao2021electric,Cao2021}. It was originally believed that the superconductivity and the correlated insulator might have the same origin, reminiscent of the cuprate phase diagram. However, this scenario is challenged by further experiments showing that superconductivity is more robust \cite{Lu2019,Saito2020independent,Stepanov2020untying,Liu2021tuning}, i.e., it can exist without any nearby correlated insulating states. Therefore, one might wonder if superconductivity and correlated states can come from completely different origins \cite{Chou2019}. 

The recent observation of ferromagnetism \cite{Zhou2021} and superconductivity \cite{Zhou2021_SC_RTG} in the moir\'eless ABC-stacked rhombohedral trilayer graphene (RTG) provides a new perspective to the origin of superconductivity in the graphene based systems. There are two distinct superconducting regions which are coined SC1 and SC2. The SC1 phase emerges from a paramagnetic normal state and is consistent with the Pauli-limited spin-singlet superconductivity. On the other hand, the SC2 phase arises from a spin-polarized, valley-unpolarized half metal and is insensitive to an applied in-plane magnetic field, implying a non-spin-singlet pairing. The moir\'eless RTG exhibits a number of phenomena (e.g, flavor polarization and superconductivity) that have been seen in the magic-angle twisted bilayer graphene. It is conjectured that the superconductivity observed in all graphene systems has the same origin \cite{Zhou2021_SC_RTG}.

We investigate the electron-acoustic-phonon coupling as a candidate mechanism for the superconductivity in RTG. We show that the leading pairings are the $s$-wave spin-singlet and $f$-wave spin-triplet pairings, which yield the same $T_c$ in our model, because the acoustic-phonon-mediated attraction respects an enlarged  SU(2)$\times$SU(2) symmetry, i.e., independent spin rotational SU(2) symmetry within each valley. We find that superconductivity prevails in a wide range of doping with $T_c\sim 1$K, arises that acoustic-phonon-mediated superconductivity is quite likely. We discuss how to understand the experimental results within the acoustic-phonon-mediated superconductivity scenario. Specifically, the SC1 phase (paramagnetic normal state) can be explained by the $s$-wave spin-singlet, and the SC2 phase (ferromagnetic normal state) can be explained by the $f$-wave equal-spin pairing. Both pairing symmetries are allowed by the electron-acoustic-phonon coupling. Our results suggest that the superconductivity in RTG very likely to be induced by the electron-acoustic-phonon coupling. 

\begin{figure}[t!]
	\includegraphics[width=0.4\textwidth]{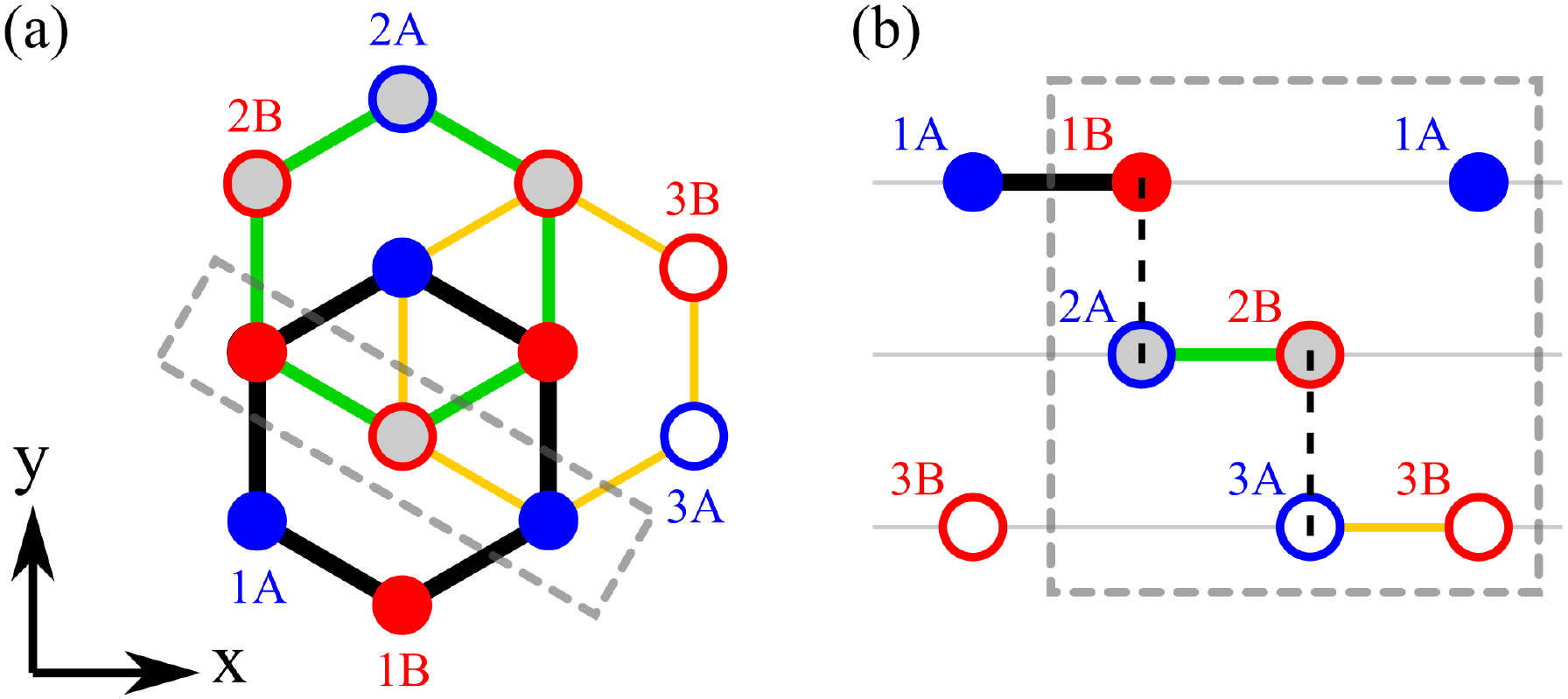}
	\caption{Lattice structure of rhombohedral trilayer graphene. (a) RTG in the $xy$ plane. For each layer, we illustrate a hexagon to specify the relative position in the $xy$ plane. 1A, 2A, and 3A (1B, 2B, and 3B) denote the sublattice A (B) in the layer 1, 2, and 3 respectively. (b) The cross section view. At K and $-$K points, the intra-layer hybridizations can be ignored, and the nearest neighbor inter-layer couplings generate dimerization in 1B-2A and 2B-3A bonds (black dashed bonds). 1A and 3B sites are the low-energy sites in this simplified picture.}
	\label{Fig:ABC}
\end{figure}

\textit{Single-particle model. --} The RTG consists of three layers of graphene with a particular ABC stacking pattern as illustrated in Fig. \ref{Fig:ABC}. 
To describe the single-particle bands near the K and $-$K valleys, we use a $\vex{k}\cdot\vex{p}$ Hamiltonian \cite{Zhang2010,Ho2016} given by   
\begin{align}\label{Eq:H_lowE}
	\hat{H}_0=\sum_{\vex{k}}\hat{\Psi}^{\dagger}(\vex{k})\hat{h}(\vex{k})\hat{1}_{s}\hat{\Psi}(\vex{k}),
\end{align}
where $\hat{h}(\vex{k})=\hat{h}_+(\vex{k})\oplus \hat{h}_-(\vex{k})$, $\hat{h}_{\pm}(\vex{k})$ is a 6-by-6 matrix associated with the low-energy Hamiltonian near $\pm$K valley, $\hat{1}_{s}$ is the identity matrix in the spin space, and $\hat{\Psi}(
\vex{k})$ is a 24-component column vector made of the fermionic annihilation operator $\psi_{\tau\sigma l s}$ with sublattice $\sigma$, spin $s$, layer $l$, and valley $\tau$. A detailed account of the $\hat{h}(\vex{k})$ is provided in the supplemental material \cite{SM}.

The RTG can be viewed as a generalized Dirac system with large chirality in the low-energy limit \cite{Zhang2010}. The low-energy states have large probabilities on the 1A and 3B sites, motivating an effective 2-by-2 Hamiltonian per spin per valley \cite{Zhang2010,Ho2016}. As illustrated in Fig.~\ref{Fig:ABC}(b), the 1A and 3B sites do not have adjacent atoms in the second layer. At the K and $-$K points, the couping between two sublattices within one layer effectively vanishes due to the threefold rotational symmetry, and the inter-layer nearest neighbor tunnelings tend to form dimerized bonds in 1B-2A and 2B-3A bonds, leaving out 1A and 3B sites. Because at least one of the sublattices in each layer is pushed to high energy by the dimerization, superconducting states with intra-layer inter-sublattice pairing structures should be suppressed for RTG with low carrier density.

The single-particle Hamiltonian in Eq.~(\ref{Eq:H_lowE}) can be diagonalized in the band basis as follows:
\begin{align}
	\hat{H}_0=\sum_{b=1}^6\sum_{\tau=\pm}\sum_{s=\uparrow,\downarrow}\epsilon_{\tau,b}(\vex{k})c^{\dagger}_{\tau b s}(\vex{k})c_{\tau b s}(\vex{k}),
\end{align}
where $\epsilon_{\tau,b}(\vex{k})$ encodes the energy-momentum dispersion of $b$th band and valley $\tau$K, and $c_{\tau b s}(\vex{k})$ is a fermionic annihilation operator of $b$th band, valley $\tau$K, and spin $s$. The operators in physical basis $\psi_{\tau\sigma l s}$ and the operators in band basis $c_{\tau b s}$ are connected by $\psi_{\tau\sigma l s}(\vex{k})=\sum_b\Phi_{\tau b, \sigma l}(\vex{k})c_{\tau b s}(\vex{k})$, where $\Phi_{\tau b, \sigma l}(\vex{k})$ is the wavefunction of band $b$ with valley $\tau$K. In addition, the (spinless) time-reversal symmetry provides constraints: $\epsilon_{+,b}(\vex{k})=\epsilon_{-,b}(-\vex{k})$ and $\Phi_{+ b, \sigma l}(\vex{k})=\Phi_{- b, \sigma l}^*(-\vex{k})$. 

\begin{figure}[t!]
	\includegraphics[width=0.45\textwidth]{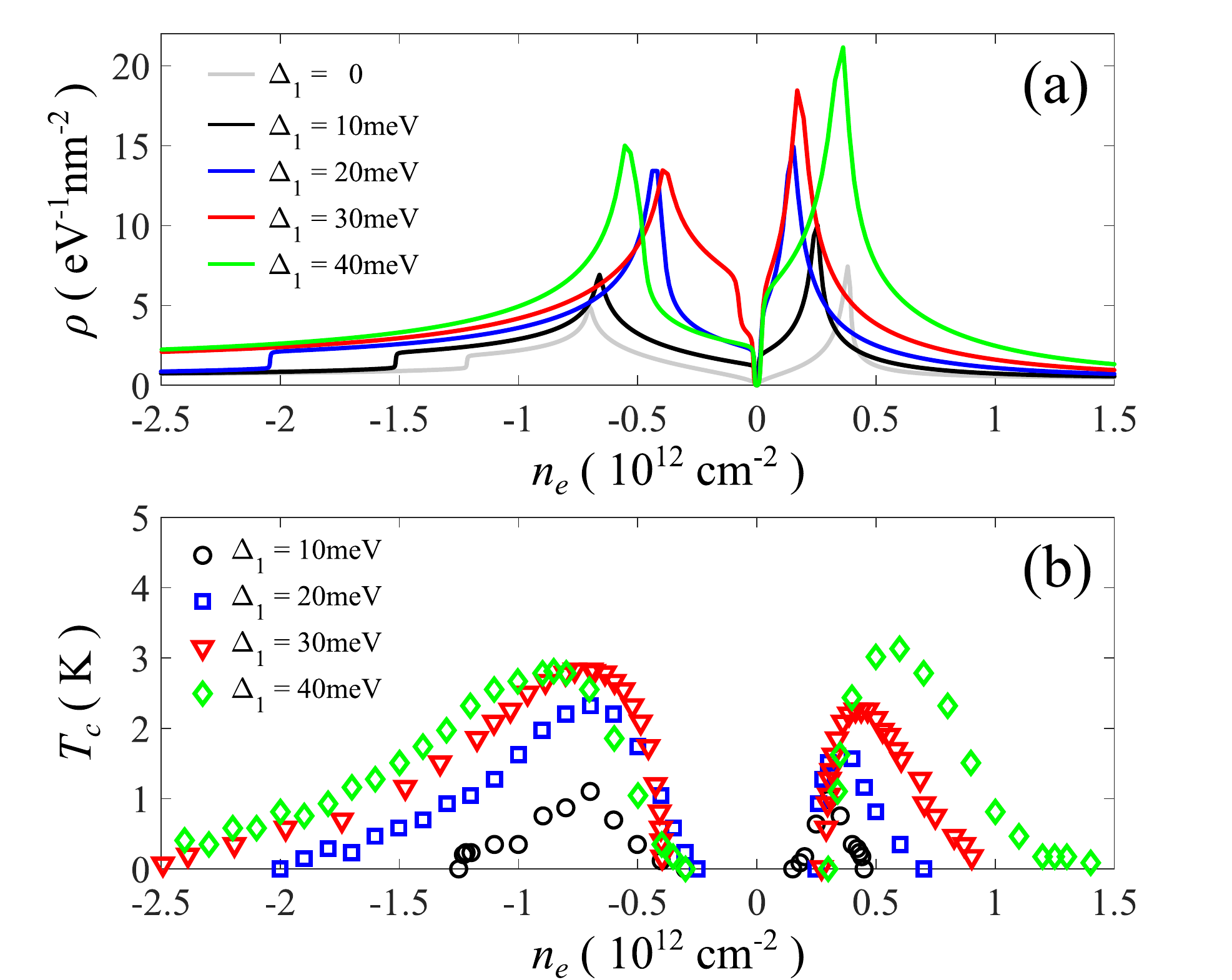}
	\caption{Total density of states ($\rho$) and superconducting transition temperature ($T_c$). $n_e$ is the total doping density, and the doping density per flavor is $n_e/4$ as no flavor polarization is considered in our model. (a) $\rho$ as a function of the $n_e$. $\Delta_1$ corresponds to the out-of-plane displacement field. A larger $\Delta_1$ can enhance the peaks at VHS. The numerical results are obtained by computing a $4001\times 4001$ momentum grid. (b) $T_c$ for acoustic-phonon-mediated superconductivity as a function of $n_e$. The peaks of $T_c$ trace roughly the positions of VHS, and the largest $T_c$ is about 3K. Moving away from the VHS, we find sizable superconducting regimes for $\Delta_1\ge 20$meV. The $T_c$ is extracted by solving Eqs.~(\ref{Eq:LGE_1}) and (\ref{Eq:LGE_2}) with a $71\times 71$ momentum grid. The cutoff of the wavevector $\Lambda\approx 0.45$nm$^{-1}$ is used in all the plots.}
	\label{Fig:DOS_Tc}
\end{figure}

We use the same model parameters as in Ref.~\cite{Zhou2021} and compute the band structures. The RTG low-energy bands (i.e., the first valence band and the first conduction band near $E=0$) feature a number of interesting properties \cite{Zhang2010,Zhou2021}.
In the absence of an out-of-plane displacement field (corresponding to $\Delta_1$), each valley develops three Dirac points \cite{Zhang2010}. When three Dirac cones merge, Van Hove singularity (VHS) develops.
A finite displacement field gaps out these Dirac touching points. In addition, annular Fermi surfaces manifest in the hole doped regime. 
We plot the calculated density of states (DOS) as a function of doping density in Fig.~\ref{Fig:DOS_Tc}(a). In particular, the VHS doping can be controlled by the displacement field, corresponding to the parameter $\Delta_1$ in our calculations.

\textit{Pairing symmetry. --} In the graphene based materials, all internal degrees of freedom (i.e., spin, valley, and sublattice) play essential roles in the pairing symmetry of the superconductivity \cite{Wu2018,Wu2019_phonon,Chou2021correlation}. We consider only the inter-valley Cooper pairs. The intra-valley superconductivity \cite{Einenkel2011,Sun2021}, corresponding to a finite-momentum Cooper pair, is generically suppressed in RTG low-energy bands because $\epsilon_{\tau,b}(\vex{k})\neq\epsilon_{\tau,b}(-\vex{k})$ for each valley $\tau$.
For a pair of electrons from different valleys, the combination of $\mathcal{C}_{3z}$ (three-fold rotation about hexagon center) and spin SU(2) symmetry can classify angular momentum states associated with $|L_z|=0,1,2,3$, corresponding to $s$-, $p$-, $d$-, and $f$-wave pairings respectively.
The $s$-wave and $f$-wave pairings are intra-sublattice while the $p$-wave and $d$-wave pairings are inter-sublattice; the $s$-wave and $d$-wave pairings are spin-singlet while the $p$-wave and $f$-wave pairings are spin-triplet \cite{Wu2019_phonon,Chou2021correlation}.

In RTG, the low-energy states have large amplitudes on the 1A and 3B sites. As a result, we find that the inter-sublattice pairings within the same layer (i.e., $p$-wave and $d$-wave pairings) are strongly suppressed in the low-energy bands. Thus, we focus only on the intra-sublattice pairings, i.e., $s$-wave spin-singlet and $f$-wave spin-triplet pairings. See \cite{SM} for a brief discussion on the inter-sublattice pairing.

\textit{Acoustic-phonon-mediated superconductivity. --} The low-energy bands of RTG manifest large DOS and VHS [Fig.~\ref{Fig:DOS_Tc}(a)], allowing for interesting many body phenomena including superconductivity.
We study superconductivity mediated by the in-plane acoustic longitudinal phonon modes \cite{Wu2019_phonon,Wu2020_TDBG}. The contributions from the optical phonons are subleading because the corresponding electron-phonon couplings have intra-layer inter-sublattice structures \cite{Wu2018}, which are less effective in mediating pairings.
After integrating out the acoustic phonon and ignoring the retardation effect,
the Bardeen-Cooper-Schrieffer (BCS) pairing interaction is given by
\begin{align}
	\hat{H}_{\text{BCS,ph}}=-g_0\sum_{\sigma,\sigma',l,s,s'}\int d^{2}\vex{r}\,\psi^{\dagger}_{+\sigma l s}\psi^{\dagger}_{-\sigma' l s'}\psi_{-\sigma' l s'}\psi_{+\sigma l s},
\end{align}
where $g_0$ is the coupling constant encoding the phonon-mediated effective attraction between the electrons. Note that the electrons only interact within the same layer because the attraction is mediated by the in-plane acoustic longitudinal phonon modes. The coupling constant $g_0=D^2/(\rho_m v_s^2)$, where $D$ is the deformation potential, $\rho_m$ is the mass density of monolayer graphene, and $v_s$ is the velocity of acoustic longitudinal phonon. With $D=30$eV, $\rho_m=7.6\times 10^{-8}$g/cm$^2$ \cite{Efetov2010}, and $v_s=2\times 10^6$cm/s, we obtain $g_0\approx474$meV$\cdot$nm$^2$ \cite{Wu2020_TDBG}. We note that the value of the deformation potential $D$ is not precisely known, and it might be off by a factor of $2$ \cite{Wu2019_phonon}.

We focus only on the first conduction band (electron doping) and the first valence band (hole doping). These two bands are separated by an energy gap $\sim \Delta_1$ which is varied from 10meV to 40meV in this work. Therefore, we can adopt the single-band approximation (to where the Fermi energy $E_F$ lies).
Besides the single-band approximation, we focus on the intra-sublattice pairings (i.e., $s$-wave and $f$-wave) as the inter-sublattice pairings are suppressed energetically. The projected BCS pairing interaction (to the $b$th band) is given by 
\begin{align}
\label{Eq:H_BCS_b}\hat{H}_{\text{BCS}}'=&\frac{-1}{\mathcal{A}}\sum g_{\vex{k},\vex{k}'}^{(b)}c^{\dagger}_{+bs}(\vex{k})c^{\dagger}_{-bs'}(-\vex{k})c_{-bs'}(-\vex{k}')c_{+bs}(\vex{k}'),\\
g_{\vex{k},\vex{k}'}^{(b)}=&g_0\sum_{\sigma,l}\left|\Phi_{+,b;l,\sigma}(\vex{k})\right|^2\left|\Phi_{+,b;l,\sigma}(\vex{k}')\right|^2,
\end{align}
where $\sum\equiv \sum_{s,s'}\sum_{\vex{k},\vex{k'}}$ in Eq.~(\ref{Eq:H_BCS_b}), $\mathcal{A}$ is the area of the system, and $g_{\vex{k}.\vex{k}'}^{(b)}$ is the momentum-dependent coupling constant in the $b$th band. In the absence of Zeeman splitting, the $s$-wave spin-singlet and the $f$-wave spin-triplet pairings are described by the same interaction, and the transition temperatures are exactly degenerate. This is because that the acoustic-phonon-mediated attraction respects an enlarged SU(2)$\times$SU(2) symmetry, i.e., independent spin rotational SU(2) symmetry within each valley.
Optical phonons \cite{Wu2018} may break the degeneracy between $s$-wave and $f$-wave parings, but this effect is subleading. In addition, a sufficiently large Zeeman field suppresses all the singlet pairings, making the spin-triplet pairing the only possibility.

With the mean field approximation, $\hat{H}_0+\hat{H}_{\text{BCS}}'$ [given by Eqs.~(\ref{Eq:H_lowE}) and (\ref{Eq:H_BCS_b})] becomes:
\begin{align}
\nonumber\hat{H}_{\text{MFT}}=&\sum_{s,s'}\sum_{\vex{k}}\mathcal{C}^{\dagger}_{ss'}(\vex{k})\hat{h}_{\text{BdG},s's}(\vex{k})\mathcal{C}_{ss'}(\vex{k})\\
&+\mathcal{A}\sum_{s,s'}\sum_{\vex{k},\vex{k'}}\Delta^*_{s's}(\vex{k})\left[\left(g^{(b)}\right)^{-1}\right]_{\vex{k},\vex{k}'}\Delta_{s's}(\vex{k}'),
\end{align}
where
\begin{align}
\mathcal{C}^{T}_{ss'}(\vex{k})=&[c_{+ bs}(\vex{k}); c_{-bs'}^{\dagger}(-\vex{k})],\\
\hat{h}_{\text{BdG},s's}=&\left[\begin{array}{cc}
	\epsilon_{+}(\vex{k})-E_F & \Delta_{s's}(\vex{k})\\[2mm]
	\Delta^*_{s's}(\vex{k}) & -\epsilon_{-}(-\vex{k})+E_F
\end{array}\right],\\
\Delta_{s's}(\vex{k}')=&\frac{1}{\mathcal{A}}\sum_b\sum_{\vex{k}'}g_{\vex{k},\vex{k}'}^{(b)}\left\langle c_{-bs'}(-\vex{k}')c_{+bs}(\vex{k}')\right\rangle.
\end{align}
To extract the transition temperature, we treat $\Delta_{ss'}$ to be infinitesimal and derive the linearized gap equation \cite{Wu2020_TDBG} (see \cite{SM} for a derivation) as follows:
\begin{align}
\label{Eq:LGE_1}\Delta_{s's}(\vex{k})=&\sum_{\vex{k}'}\chi_{\vex{k},\vex{k}'}\Delta_{s's}(\vex{k}'),\\
\label{Eq:LGE_2}\chi_{\vex{k},\vex{k}'}=&\frac{g_{\vex{k},\vex{k}'}^{(b)}}{\mathcal{A}}\frac{\tanh\left[\frac{\epsilon_{+b}(\vex{k}')-E_F}{2k_BT}\right]}{2\epsilon_{+b}(\vex{k}')-2E_F}.
\end{align}
Notice that Eq.~(\ref{Eq:LGE_1}) is a self-consistent eigenvalue problem with the discrete wavevectors being the indices of matrix. The transition temperature $T_c$ is determined by the highest $T$ such that $\chi_{\vex{k},\vex{k}'}$ yields an eigenvalue $1$. 

We numerically solve Eqs.~(\ref{Eq:LGE_1}) and (\ref{Eq:LGE_2}) and plot $T_c$ as a function of doping in Fig~\ref{Fig:DOS_Tc}(b). Again, the $s$-wave spin-singlet and the $f$-wave spin-triplet pairings yield the same $T_c$. We find that superconductivity prevails with $T_c$ peaked at the VHS doping in both the electron doping and the hole doping for $\Delta_1=10-40$meV. As shown in Fig.~\ref{Fig:DOS_Tc}(b), $T_c$ can be of order 1K for doping densities away from VHS doping, so it is not that superconductivity manifests only at the VHS (see Ref.~\cite{Lothman2017} for a similar finding). Technically, the prevalence of superconductivity arises from the energy dependence of $\chi_{\vex{k},\vex{k}'}$ in Eq.~(\ref{Eq:LGE_2}), where energy levels away from $E_F$ can still contribute. We also check the inter-sublattice pairings ($p$-wave and $d$-wave) and confirm that the associated $T_c$ is too small to be resolved in our numerical calculations. This is consistent with the physical intuition that the inter-sublattice pairings within the same layer are energetically suppressed. Thus, we conclude that acoustic-phonon mediated superconductivity is quite probable in RTG, and the pairing symmetry is either $s$-wave spin-singlet or $f$-wave spin-triplet.

So far, we use the bare electron-phonon coupling to estimate the superconductivity mediated by the acoustic phonon. The Coulomb repulsion can reduce the effective attraction and suppress the superconductivity. However, such a suppression is usually addressed by the retardation effect of the electron-phonon coupling---the so-called $\mu^*$ effect \cite{Coleman2015introduction,MorelAnderson1962}. Thus, it is possible that the $g_0$ here is over estimated, and the $T_c$ for the actual system might be smaller.
It is reasonable to expect that $\mu^*$ is likely to be small because the large DOS can lead to a strong screening of Coulomb interaction.

\begin{figure}[t!]
	\includegraphics[width=0.45\textwidth]{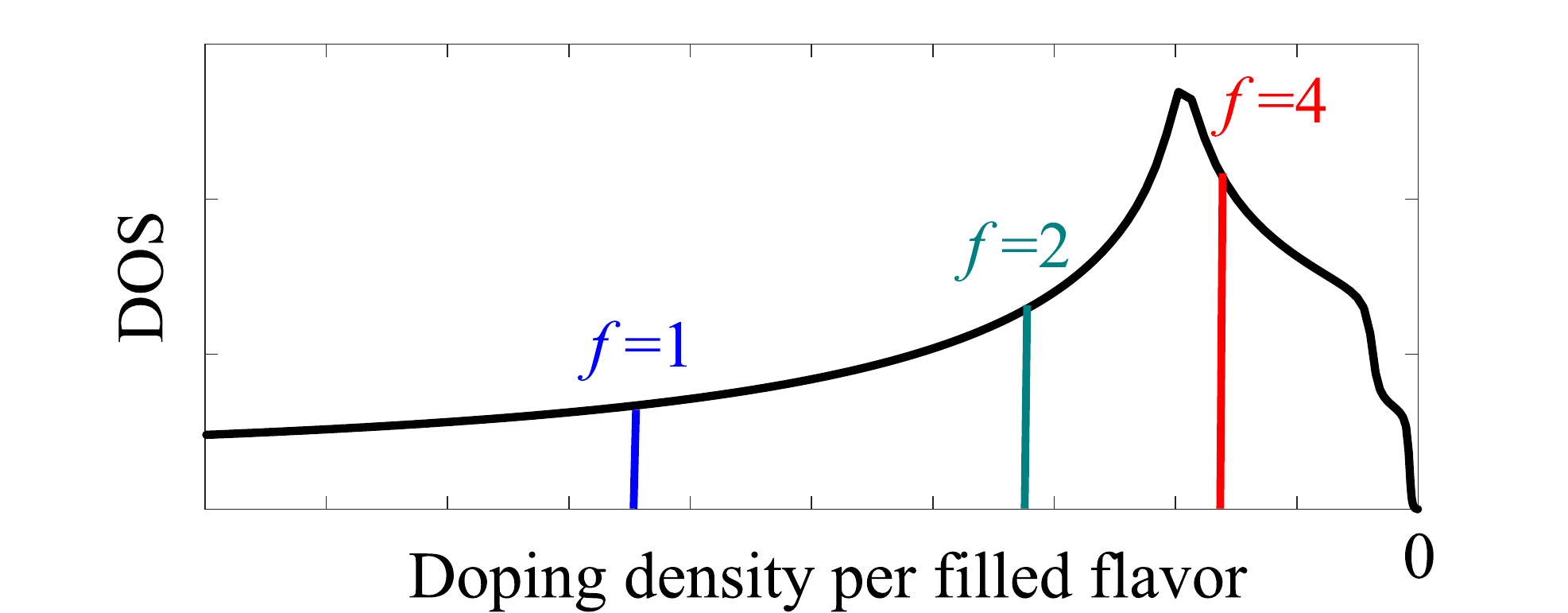}
	\caption{Density of states and flavor polarization. We consider three different states with the same total density. $f$ denotes the number of filled flavor (spin and valley). $f=1$ indicates a spin-polarized valley-polarized state; $f=2$ indicates a spin-polarized valley-unpolarized state; $f=4$ indicates no flavor polarization. The DOS depends on both the flavor polarization and the doped density. In this illustration, we use $\Delta_1=30$meV and focus on the hole doping side.
	}
	\label{Fig:DOS_flavor}
\end{figure}

\textit{Discussion. --} In the RTG experiment \cite{Zhou2021_SC_RTG}, there are two distinct superconducting regions: SC1 and SC2. The former is consistent with a Pauli-limited $s$-wave spin-singlet pairing superconductor, and the latter is most likely a non-spin-singlet pairing superconductor. Both SC1 and SC2 superconducting states can be explained by the electron-acoustic-phonon coupling, which allows for $s$-wave spin-singlet and $f$-wave spin-triplet pairings. In a paramagnetic normal state (corresponding to SC1), $s$-wave is generically favored because the subleading pairing mechanisms (such as optical phonon \cite{Wu2018}) generically enhance the $s$-wave channel. In a ferromagnetic normal state (corresponding to SC2), pairing among different spins (such as spin-singlet pairing) is suppressed, but the $f$-wave equal-spin pairing may survive. Thus, electron-acoustic-phonon coupling can produce the same phenomenology in both SC1 and SC2 regions.

To understand superconductivity in RTG experiments \cite{Zhou2021_SC_RTG}, we need to take into account the flavor polarization \cite{Zhou2021}. 
The flavor polarization induces a half metal (spin-polarized valley-unpolarized state) and a quarter metal (spin-polarized valley-polarized state) \cite{Zhou2021}. As illustrated in Fig.~\ref{Fig:DOS_flavor}, the DOS depends not only on the doping density but also on the flavor polarization pattern. In general, the flavor polarized state, corresponding to higher doping per filled flavor, is energetically favored over the unpolarized states at a doping density that would have VHS in the noninteracting model. These complications impose further restrictions for superconductivity. As discussed in Ref.~\cite{Zhou2021_SC_RTG}, the presence of flavor polarization is not detrimental to the phonon-mediated superconductivity scenario. The Stoner criteria depends essentially on the DOS at the Fermi level, while the BCS superconductivity has a very different dependence in DOS [see Eqs.~(\ref{Eq:LGE_1}) and (\ref{Eq:LGE_2})]. In Fig.~\ref{Fig:DOS_Tc}, we show that the value of $T_c$ can still be of order 1K away from the VHS doping, suggesting that one should observe superconductivity in the regimes where other competing orders are absent. Using $\Delta_1=30$meV and $n_e=-2.0\times 10^{12}$cm$^{-2}$ (corresponding to $n_e=-0.5\times 10^{12}$cm$^{-2}$ per flavor), we obtain $T_c\sim 0.5$K for the acoustic-phonon-mediated superconductivity. The $T_c$ value is off by a factor of 5 compared with the $T_c=100$mK observed in SC1 \cite{Zhou2021_SC_RTG}. With respect to SC2, the experimental density is $-0.5\times 10^{12}$cm$^{-2}$ (corresponding to $n_e=-0.25\times 10^{12}$cm$^{-2}$ per flavor). With $\Delta_1=20$meV, the estimated $T_c$ for SC2 is 1.6K which is 30 times larger than the experimental estimate $\sim 50$mK \cite{Zhou2021_SC_RTG}. The low experimental $T_c$ is possibly accounted for by the Coulomb repulsion, i.e., the $\mu^*$ effect (for both SC1 and SC2), by impurity scattering (primarily for SC2), and by the magnon fluctuation (for SC2), and the mean field nature of our theory may overestimate the $T_c$ in two dimensions. The difference in $T_c$ between SC1 and SC2 might be due to the optical phonon contribution, which favors $s$-wave spin-singlet pairing.
Moreover, inter-valley scattering caused by atomic-scale lattice defects and scattering from the edge can also suppress the $T_c$ for the unconventional spin-triplet phase SC2. In the latter case, the SC2 phase would be stabilized with increasing the sample size. 
Our results show that acoustic phonons produce superconductivity in RTG, but a more quantitative understanding must await more experimental data in more samples and future theoretical investigations.
Using our current theory and the electron-phonon coupling estimates \cite{Min2011}, we predict a robust existence of superconductivity in ABCA-stacked graphene as well.

One interesting prediction based on our theory is that a sufficiently large Zeeman field can destroy the $s$-wave spin-singlet pairing, and then the $f$-wave equal-spin pairing becomes the leading superconducting instability. 
However, such a novel superconductor-superconductor transition has not been confirmed experimentally. In the RTG experiment \cite{Zhou2021_SC_RTG}, SC1 is suppressed by a Zeeman field, and no sign of re-entrant superconductivity is reported. Although acoustic phonons give rise to the same $T_c$ for the $s$-wave and $f$-wave pairings, the $T_c$ for the $f$-wave spin-triplet pairing may be reduced due disorder scattering in the bulk and inter-valley scattering from the sample boundary. In addition, the subleading optical phonon contribution can lift the degeneracy between $s$-wave and $f$-wave pairings, and $f$-wave pairing  has a lower $T_c$ typically.
Therefore, it is possible that the resulting $T_c$ for the $f$-wave equal-spin pairing is too small to be detectable, but more systematic investigations at lower temperatures are required.
Based on our conclusion about RTG $s$-wave (SC1) and $f$-wave (SC2) pairings, we predict SC1 (SC2) phase to be robust (vulnerable) to increasing disorder in the system.

Since acoustic phonons are important, one should see a linear-in-$T$ resistivity at higher temperatures ($T>T_{\text{BG}}/4$ \cite{Hwang2008,Min2011,Wu2019_phonon}, where $T_{\text{BG}}$ is the Bloch-Gr\"uneisen temperature.), depending on the doping, but we estimate it to be above 10K-20K \cite{Hwang2008,Min2011}, and the electron-phonon coupling parameter extracted from such a linear-in-$T$ resistivity should have approximate consistency with the observed $T_c$ \cite{Min2011,Wu2019_phonon,Li2020,Polshyn2019,Cao2020PRL,Chu2021phonons}. The same is true for spin or valley fluctuation mediated SC too. In the RTG experiment \cite{Zhou2021_SC_RTG}, a linear-in-$T$ resistivity is not seen for $T\le 20$K, but we predict that there should be a phonon-induced linear-in-$T$ resistivity for $T>20$K above the superconducting state.

We comment on alternative mechanisms for superconductivity. Owing to the presence of ferromagnetism, it is natural to speculate that spin fluctuations might play an important role \cite{Chou2021correlation,Fischer2021unconventional,Wang2021}. However, the spin-fluctuation mechanism is not consistent with either SC1 or SC2. First of all, superconductivity driven by the ferromagnetic spin fluctuations is spin-triplet pairing \cite{Chou2021correlation}, which cannot account for SC1. In addition, the normal state of the spin-fluctuation-induced superconductivity \cite{Chou2021correlation} cannot be fully spin polarized because the paramagnon fluctuation, which provides spin-triplet pairing, is negligible. Fluctuations of valley degrees of freedom \cite{Wang2021} may still generate spin-triplet superconductivity. Nevertheless, a microscopic justification of such a mechanism is not clear. Our acoustic-phonon-mediated superconductivity can explain both SC1 and SC2. It is unlikely but not impossible that SC1 and SC2 have different pairing mechanisms. 

The observed superconductivity and ferromagnetism in RTG are reminiscent of that in various graphene moir\'e systems. 
We establish that the superconductivity in RTG can be explained by the electron-acoustic-phonon coupling, suggesting that superconductivity and ferromagnetism have different origins. This might also be true for other graphene systems, including magic-angle twisted bilayer graphene \cite{tbg1,tbg2,Yankowitz2019,Polshyn2019,Cao2020PRL,Sharpe2019,Lu2019,Kerelsky2019,Jiang2019,Xie2019_spectroscopic,Choi2019,Serlin2020,Park2021flavour,Cao2020tunable,Liu2020tunable}, magic-angle twisted trilayer graphene \cite{Park2021tunable,Hao2021electric,Cao2021}, and mutlilayer rhombohedral graphene \cite{Shi2020}.

\begin{acknowledgments}
	\textit{Acknowledgments.--} We are grateful to Andrea Young for useful discussions.
	This work is supported by the Laboratory for Physical Sciences (Y.-Z.C. and S.D.S.), by JQI-NSF-PFC (supported by NSF
	grant PHY-1607611, Y.-Z.C.), and NSF DMR1555135 (CAREER, J.D.S.) 
\end{acknowledgments}	




\newpage \clearpage 

\onecolumngrid

\begin{center}
	{\large
		Acoustic-phonon-mediated superconductivity in rhombohedral trilayer graphene
		\vspace{4pt}
		\\
		SUPPLEMENTAL MATERIAL
	}
\end{center}

\setcounter{figure}{0}
\renewcommand{\thefigure}{S\arabic{figure}}
\setcounter{equation}{0}
\renewcommand{\theequation}{S\arabic{equation}}

In this supplemental material, we provide technical details of main results in the main text.

\section{Single-particle Hamiltonian}

The 6-by-6 matrix $\hat{h}_{\tau}(\vex{k})$ is given by \cite{Zhang2010,Zhou2021}
\begin{align}\label{Eq:h_k}
	\hat{h}_{\tau}(\vex{k})=\left[\begin{array}{cccccc}
		\Delta_1+\Delta_2+\delta& \frac{1}{2}\gamma_2 & v_0\Pi^{\dagger}_{\vex{k}} & v_4\Pi^{\dagger}_{\vex{k}} & v_3\Pi_{\vex{k}} & 0 \\[2mm]
		\frac{1}{2}\gamma_2 & \Delta_2-\Delta_1+\delta & 0 & v_3\Pi^{\dagger}_{\vex{k}} & v_4\Pi_{\vex{k}} & v_0\Pi_{\vex{k}} \\[2mm]
		v_0\Pi_{\vex{k}} & 0 & \Delta_1+\Delta_2 & \gamma_1 & v_4\Pi^{\dagger}_{\vex{k}} & 0 \\[2mm]
		v_4\Pi_{\vex{k}} & v_3\Pi_{\vex{k}} & \gamma_1 & -2\Delta_2 & v_0\Pi^{\dagger}_{\vex{k}} & v_4\Pi^{\dagger}_{\vex{k}} \\[2mm]
		v_3\Pi^{\dagger}_{\vex{k}} & v_4\Pi^{\dagger}_{\vex{k}} & v_4\Pi_{\vex{k}} & v_0\Pi_{\vex{k}} & -2\Delta_2 & \gamma_1 \\[2mm]
		0 & v_0\Pi^{\dagger}_{\vex{k}} & 0 & v_4\Pi_{\vex{k}} & \gamma_1 & \Delta_2-\Delta_1
	\end{array}
	\right],
\end{align}
where $\Pi_{\vex{k}}=\tau k_x+ik_y$ ($\tau= 1,-1$ for valleys K and $-$K respectively), $v_j=\sqrt{3}a_0\gamma_j$, $\gamma_j$ is the bare hopping matrix element, and $a_0=0.246$nm is the lattice constant of graphene. The basis of $\hat{h}_{\tau}(\vex{k})$ is (1A,3B,1B,2A,2B,3A). Note that the first two elements, 1A and 3B, are the low-energy sites as discussed in the main text.\\

We use the same parameters in Ref.~\cite{Zhou2021}. Specifically, $\gamma_0=3.1$eV, $\gamma_1=0.38$eV, $\gamma_2=-0.015$eV, $\gamma_3=-0.29$eV, $\gamma_4=-0.141$eV, $\delta=-0.0105$eV, and $\Delta_2=-0.0023$eV. The value of $\Delta_1$ corresponds to the out-of-plane displacement field, and we vary it from 10 to 40meV.

\section{Inter-sublattice pairing}

In the main text, we focus only on the intra-sublattice pairing, corresponding to $s$-wave and $f$-wave, because the inter-sublattice pairings within a layer have higher energies in the system. For completeness, we provide the projected interactions in the inter-sublattice pairing as follows:
\begin{align}
	\hat{H}_{\text{BCS}}'=&-\frac{1}{\mathcal{A}}\sum_{\vex{k},\vex{k'}}J_{\vex{k},\vex{k}'}^{(b)}c^{\dagger}_{+,\uparrow}(\vex{k})c^{\dagger}_{-,\uparrow}(-\vex{k})c_{-,\uparrow}(-\vex{k}')c_{+,\uparrow}(\vex{k}'),\\
	\nonumber J_{\vex{k},\vex{k}'}^{(b)}=&g\sum_{l}\Phi_{+,b;l,A}^*(\vex{k})\Phi_{+,b;l,B}(\vex{k})\Phi_{+,b;l,B}^*(\vex{k}')\Phi_{+,b;l,A}(\vex{k}')\\
	&+g\sum_{l}\Phi_{+,b;l,B}^*(\vex{k})\Phi_{+,b;l,A}(\vex{k})\Phi_{+,b;l,A}^*(\vex{k}')\Phi_{+,b;l,B}(\vex{k}'),
\end{align}
where $J_{\vex{k},\vex{k}'}^{(b)}$ is the effective inter-sublattice pairing interaction projected to band $b$. With the expression of $J_{\vex{k},\vex{k}'}^{(b)}$, one can extract $T_c$ for the inter-sublattice pairing by replacing the effective pairing interaction in the gap equation. We find that the $T_c$ for inter-sublattice pairings is too small to be resolved numerically, consistent with our intuitive understanding that inter-sublattice pairings are not favorable energetically.

\section{Derivation of linearized gap equation}

\begin{figure}[t!]
	\includegraphics[width=0.7\textwidth]{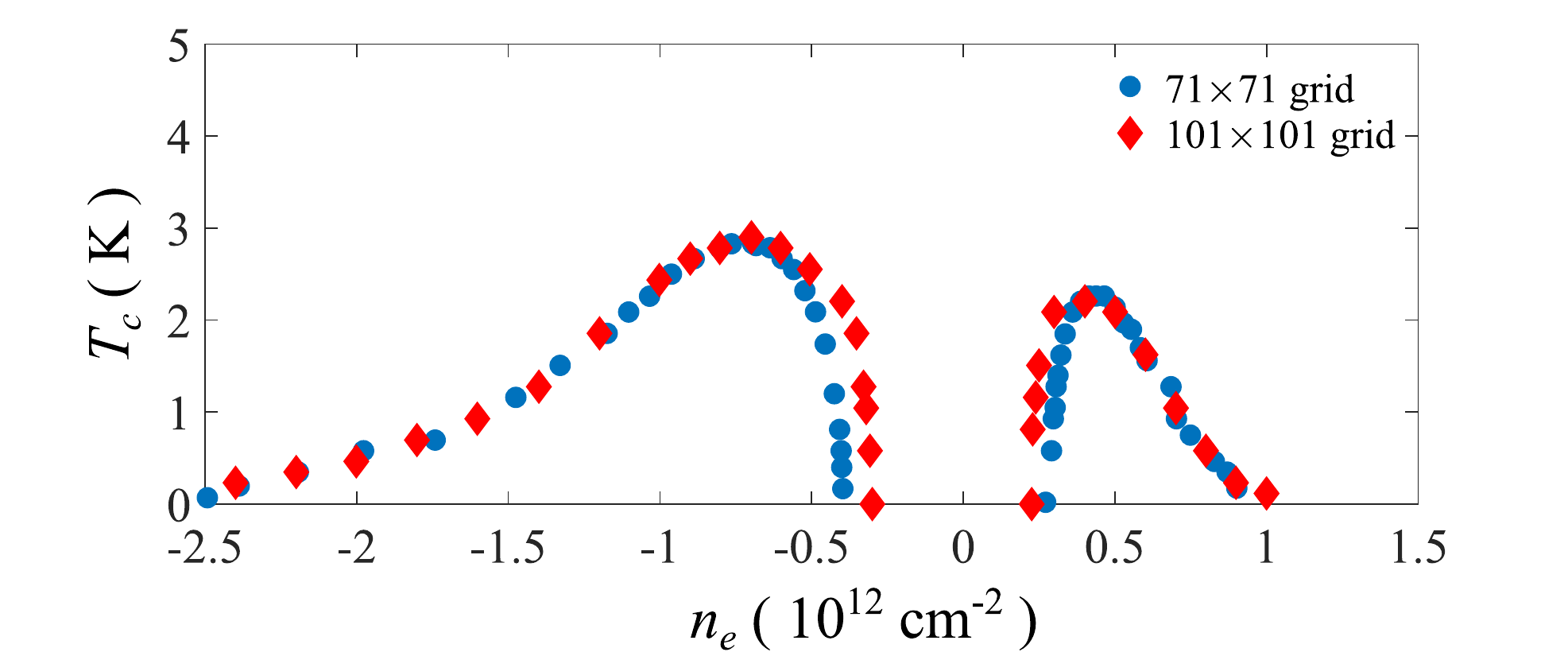}
	\caption{Superconducting transition temperature ($T_c$) as functions of doping density. We choose $\Delta_1=30$meV. The blue dots are calculated with a system size $L=2000a_0$ and a 71-by-71 momentum grid; The red diamonds are calculated with a system size $L=3000a_0$ and a 101-by-101 momentum grid. The numerical results for two sizes are very close, suggesting weak finite size effect in our numerical calculations.}
	\label{Fig:Finite_size}
\end{figure}

In the imaginary-time path integral, we integrate out the fermions and derive the effective action given by
\begin{align}
	\mathcal{S}_{\text{eff}}=&-\sum_{\omega_n,\vex{k}}\ln\left[\left(-i\omega_n+\epsilon_{+}(\vex{k})-\mu\right)\left(-i\omega_n-\epsilon_{-}(-\vex{k})+\mu\right)-|\Delta(\vex{k})|^2\right]+\mathcal{A}\beta\sum_{\vex{k},\vex{k'}}\Delta^*(\vex{k})\left(g^{-1}\right)_{\vex{k},\vex{k}'}\Delta(\vex{k}'),
\end{align}
where we have suppressed the spin indices and band index for simplicity.
Near the transition temperature, we assume the $\Delta(\vex{k})$ is infinitesimal and expand the logarithm. As a result, we obtain the Landau free energy density as follows:
\begin{align}
	\mathcal{F}=&\frac{\mathcal{S}_{\text{eff}}}{\beta \mathcal{A}}\approx\text{const}+\frac{1}{\beta\mathcal{A}}\sum_{\omega_n,\vex{k}}\frac{|\Delta(\vex{k})|^2}{\left(-i\omega_n+\epsilon_{+}(\vex{k})-\mu\right)\left(-i\omega_n-\epsilon_{-}(-\vex{k})+\mu\right)}+\sum_{\vex{k},\vex{k'}}\Delta^*(\vex{k})\left(g^{-1}\right)_{\vex{k},\vex{k}'}\Delta(\vex{k}')+\mathcal{O}(|\Delta(\vex{k})|^4)\\
	=&\text{const}-\frac{1}{\mathcal{A}}\sum_{\vex{k}}\frac{1-2f(\epsilon_+(\vex{k})-\mu)}{2\left[\epsilon_+(\vex{k})-\mu\right]}|\Delta(\vex{k})|^2+\sum_{\vex{k},\vex{k'}}\Delta^*(\vex{k})\left(g^{-1}\right)_{\vex{k},\vex{k}'}\Delta(\vex{k}')+\mathcal{O}(|\Delta(\vex{k})|^4),
\end{align}
where $f(x)$ is the Fermi distribution function. We have used the spinless time-reversal symmetry yielding $\epsilon_{+}(\vex{k})=\epsilon_{-}(-\vex{k})$.
The linearized gap equation can be obtained by differentiating $\Delta^*(\vex{k})$ on $\mathcal{F}$,
\begin{align}
	&\frac{\delta \mathcal{F}}{\delta \Delta^*(\vex{k})}=0=-\frac{1}{\mathcal{A}}\frac{1-2f(\epsilon_+(\vex{k})-\mu)}{2\left[\epsilon_+(\vex{k})-\mu\right]}\Delta(\vex{k})+\sum_{\vex{k}'}\left(g^{-1}\right)_{\vex{k},\vex{k}'}\Delta(\vex{k}').
\end{align}
Note that the momentum indices can be viewed as matrix indices. After simply algebraic manipulation, we obtain
\begin{align}
	\Delta(\vex{k})=&\sum_{\vex{k}'}\chi_{\vex{k},\vex{k}'}\Delta(\vex{k}'),\\
	\chi_{\vex{k},\vex{k}'}=&\frac{g_{\vex{k},\vex{k}'}}{\mathcal{A}}\frac{1-2f(\epsilon_+(\vex{k}')-\mu)}{2\left[\epsilon_+(\vex{k}')-\mu\right]}=\frac{g_{\vex{k},\vex{k}'}}{\mathcal{A}}\frac{\tanh\left[(\epsilon_+(\vex{k}')-\mu)/(2T)\right]}{2(\epsilon_+(\vex{k}')-\mu)},
\end{align}
where $T$ is the temperature.
$\chi_{\vex{k},\vex{k}'}$ can be viewed as a matrix with label $\vex{k}$ and $\vex{k}'$. The transition temperature is obtained when $\chi_{\vex{k},\vex{k}'}$ yields an egienvalue $1$. In Fig.~\ref{Fig:Finite_size}, we show numerical results for $T_c$ with two different sizes of momentum grids. The results suggest that the finite size effect is not significant.

\end{document}